# Game-theoretic dynamic investment model with incomplete information: futures contracts

Oleg Malafeyev[*][1], Shulga Andrey[†][2]

[1]Saint-Petersburg State University, Russia

Abstract.

Over the past few years, the futures market has been successfully developing in the North-West region. Futures markets are one of the most effective and liquid-visible trading mechanisms. A large number of buyers are forced to compete with each other and raise their prices. A large number of sellers make them reduce prices. Thus, the gap between the prices of offers of buyers and sellers is reduced due to high competition, and this is a good criterion for the liquidity of the market. This high degree of liquidity contributed to the fact that futures trading took such an important role in commerce and finance. A multi-step, non-cooperative n persons game is formalized and studied

## 1 Overview

Futures are traded on specialized futures exchanges, where fair pricing is provided due to the high concentration of supply and demand. The future prices for grain, metal, gasoline, the currency that the futures exchange generates is a sure guide for any enterprise planning its activity in the market conditions. The system of financial guarantees operating on the futures exchange ensures unconditional fulfillment of obligations, reliably protects the future price of the goods.

___________________

[*]malafeyevoa@mail.ru
[†]head1.private@gmail.com

That is why all over the world banks, investment funds, wholesalers, industrial and agricultural enterprises come to the futures exchange, which gives them the opportunity not only to reduce commercial risk, but also to make a profit when trading in futures.

Futures exchanges perform specific functions for the economy: transfer risk, identify prices, increase liquidity and efficiency, increase the flow of information. Futures allow us to agree today on the price at which the purchase or sale will be made in the future. The basis is the ability to set today the price at which the future will be a purchase or sale.

*A futures contract* - is an agreement between two parties to deliver goods of a certain quantity and quality in a certain place and at a certain time in the future, at the price agreed upon today, concluded according to the rules of the exchange. The price at which a futures contract is concluded is determined by free competition among trading participants in the exchange's operating floor.
Each futures contract has two sides: the buyer and the seller. The buyer of a futures contract is called a long-stakes party, and the seller is a party with a short position. The buyer undertakes to make a purchase at a predetermined time. The seller undertakes to sell at a predetermined time. These obligations are determined by the name of the asset, the size of the asset, the term of the futures and the price agreed today. Only a whole number of futures can participate in trading.

During the term of the contract, its price depends on the state of the conjuncture for the relevant product. Buyers benefit from higher prices, as they will be able to receive the goods at a price lower than the current one. Sellers benefit from falling prices, as they contracted at a price higher than the current one. Simultaneously with the price fluctuation, the value of the contract also changes. For the holder of a long position, profits arise when prices rise, which increases the value of his contract. The fall in prices and, accordingly, the decrease in the value of the contract gives a profit to the holder of a short position. The difference in the value of a contract for a long or short position is defined as the difference between the transaction price and the current quote multiplied by the contract unit.

Each holder of the short and long positions of the contract is obligated to provide his broker with a certain amount of money as a guarantee of the performance of the contract. This deposit is called a margin. Its purpose is to protect the seller from non-fulfillment of the contract by the buyer, if the prices have fallen, and the buyer, if prices have increased. The broker uses margin to cover the losses of the client, if his contract suffers losses.

In futures trading, there are two types of margin:
*The initial margin* - is the deposit that is made when opening the futures position;
*Variation Margin* - the transfer of funds, ensuring that the cost of securing the new value of the contract follows the change in prices.

The initial margin, set in value terms, is usually 2-10% of the value of the contract. In periods of high price volatility and high risk, the exchange can set a margin at the upper limit of 10% and even higher. The margin is not the value of a futures trading operation. The money that the client transfers to a special account is his property until, as a result of some unsuccessful operation, he does not lose it. At the time of the expiration of the futures contract, the margin currently available in the account with the broker is paid to the customer back.

After the opening of the position and the introduction of the initial margin change of the futures contract price will lead to a corresponding reduction or increase in the value of the client's position. After determining the settlement price of the day, for each of the contracts and each position, the change in the value of each exchange contract is calculated based on the difference between the settlement price of the given day and the previous day or the price at which the transaction was concluded. If the situation on the futures market changes for some day in a favorable for the client side, the amount of money in the margin account increases by the size of the potential win. If these price changes are unfavorable for the client's position, then its initial margin decreases. As soon as the amount of the customer's initial margin has been reduced to the level of variation margin, additional funds may be required from the client. This requirement is caused by unfavorable changes in futures contract prices. Thus, the margin shows the profits and losses of the customer for the day. Those funds that exceed the required amount can be withdrawn by the client, but more often they remain on the account with the broker as a reserve.

Real goods can be sold and bought in two separate, but related markets - cash and futures. The prices of the cash and futures markets are very closely related. Futures markets have a solid foundation in reality.

*The cash market* - is the place where the goods change the owner for a certain price. Realization of transactions in the cash market is the purchase and sale of cash at current prices, the delivery of which is carried out immediately or within a few days after the conclusion of the transaction.

*Cash prices* - is the prices for which the goods are sold in different parts of the market. At each moment there are many cash prices de-pending on the quality of the goods, the place of delivery, the processing stage of the goods and other factors.

In the conditions of the futures market, it is a question of the future delivery of the goods in certain terms. In contrast to the cash market, the futures market has only one set of prices. The futures price is the current view of the market on how much the commodity of a certain quality will cost with certain delivery conditions at some point in the future. Time and expectations of market participants - these are the two factors that determine the difference in prices in the cash and futures markets.

Both markets exist in parallel, but at the expiration of each period, they seem to merge, destroying the existing price differences. At the time of delivery in the futures and cash markets, asset prices will be the same because of quotations for the immediate delivery of the asset in both markets. This coincidence at one point is called *convergence*. The parallel movement of these markets is due to the fact that factors leading to higher prices affect futures prices in the same direction.

For the difference between cash and futures prices, there is the concept of basis. *The basis* is the difference between the price of the goods in the cash market in a particular place and at the same price as the commodity in the futures market. The basis is calculated by subtracting the price in the futures market from the cash price, which usually means the closest futures month (period):

**Cash price - Futures price = Basis**

The basis is positive, negative and zero. If the basis is positive, then the price of the cash market exceeds the futures price or comes with a premium to the futures price. If the basis is negative, the cash price is lower than the futures price or comes at a discount to the futures price. If the basis is zero, the cash price is equal to the futures price. There can be many bases for one product at one time, because if there is only one price of a futures contract for a certain product, then there are a lot of cash prices for this product, depending on the quality and the place of delivery.

The basis is not a constant value. Since the prices of the cash and futures markets are constantly changing, the basis becomes wider or narrower. The factors that influence the size of the basis are quite a lot: they are the demand and supply for a certain period, the volume of the stock of production, the forecast for the production of goods in the current year, the supply and demand for similar products, the export and import of goods, the availability of warehouses for storage, moves and a number of others.

There are two types of dynamics of the basis:
Strengthening the basis (narrowing) is a change in the basis at which the cash price rises within a certain period of time relative to the futures price.

Weakening of the basis (expansion) - occurs when the cash price is lowered relative to the futures price for a certain period of time and becomes less stable than the futures price. The type of dynamics of the basis, most favorable for a particular market participant, depends on whether he is a seller
or the buyer. The seller wins with a strong base, and the buyer wins with a weak base.

Futures can be used in various situations: to avoid risk or to generate high returns with high risk. Futures trading can be both very risky and very profitable. Futures trading involves hedgers, speculators and arbitrageurs. The main goal of a hedger is to reduce the percentage of risk. The speculator is looking for high returns due to high risk. The purpose of the arbitrage is income without risk due to market inconsistencies.

On futures exchanges, all transactions are primarily speculative in nature or are made for the purpose of insurance against price risks.

Speculative transactions are made on the stock exchange in order to obtain profit from the purchase and sale of exchange contracts as a result of the difference between the price of the exchange contract on the day of imprisonment and the price on the day of its execution, with a favorable price change for one of the parties (the seller and the buyer). Moreover, exchange speculation is a mechanism for identifying and stabilizing prices for goods.

Futures contracts are also used in hedging. This is an operation to insure price risk by trading futures contracts. The principle of insurance here is based on the fact that if in a transaction one side loses as a seller of a real commodity, then it wins as a futures buyer for the same quantity of goods, and vice versa. The mechanism of hedging is based on the fact that the change in market prices for futures are the same in their size and direction.
There are also arbitrage transactions that are made for the purpose of making a profit due to the difference in quotations on stock exchanges in different countries.

Having sold (bought) the futures contract, the participant opens a position in the futures market. Close this position, he can either by executing a contract, or by buying (selling) the same contract. The sale and purchase of a contract of the same type compensate each other and are not taken into account in the calculations between the participants. If the futures contract sold on the exchange is not compensated by the purchase of the same contract, then with the onset of the month of execution, it must deliver a standard product. The quantity and quality of the goods supplied, the amount of payment, time, place and other conditions of delivery are strictly regulated by the rules.

Every day, customers' accounts reflect changes that have occurred in the value of contracts that they have opened. If the customer has a long position, and the prices have increased, his funds will increase, as the result for open positions increases. The gain will be the difference in the value of his open contract. Excess funds can be transferred by the client or used by him for opening new positions.

In the futures market, the transaction participant controls its investment of capital with less money than in other markets.

On the last day of the term of the futures contract, the amount of costs is zero.

## 2 Problem statement. Mathematical description of the model

In this paper, a competitive model of decision-making in the futures contract market is considered. Agents who own the initial capital, come to the futures contracts market and conclude contracts, in order to obtain maximum profit. The actions of agents are carried out at specific times. It is required to choose agents control so that the effect of investing in futures contracts is maximized.

As a method of solution we will use the method of compromise solution.

Let there be $n$ agents on the market, $i = \overline{1,n}$. They carry out their activities through brokerage offices, paying them a commission fee for their services (we will assume that the commissions are precisely defined for each contract). Let there be $s$ contracts in circulation, $j = \overline{1,s}$, acting on the time interval [0,T].

Consider the partition of the interval $0 = t_1 < t_2 < \ldots < t_f = $ T. At time $t = t_k$, $k = \overline{1, f-1}$ the market state is described by the vector of futures prices
$$x^t = (x_1^t, \ldots, x_j^t, \ldots, x_s^t),$$

where $x_j^t$ – is the futures price j – th contract at the time $t = t_k$, $k = \overline{1, f-1}$.

We will assume that at every moment $t = t_k$ agents have complete information about the history of the process. At the time $t_k$ the agent $i$ enters into a set of futures contracts.

$$u_i^t = (u_{1i}^t, u_{2i}^t, \ldots, u_{s,i}^t), \ i = \overline{1, n}, \ t \in [0, T].$$

Control $u_i^t$ is called *agent Control i* at time $t = t_k$, $k = \overline{1, f-1}$.

The set of agent controls $U^t = (u_1^t, \ldots, u_n^t)$ is called *the control of the market at time* $t = t_k$, $k = \overline{1, f-1}$.

The prices of futures contracts change at each time $t = t_k$, $k = \overline{1, f-1}$, depending on the actions of agents, on the situation prevailing in the market and from other possible control moments.
The system goes from one state to another, depending on the previous quotes and from the controls chosen by the agents according to the rule:

$$x_j^{t+1} = \beta(x_j^t, U^t),$$

where $\beta$ — is the transition operator, which is constructed on the basis of statistics collected over a sufficiently long period of time.

For simplicity, we assume that the term of the futures contract, which was concluded at the time $t$, expires at the next moment of time $t+1$.

At the end of each period of time, all futures contracts are settled, new futures contracts prices are determined and the revenue (losses) that each agent receives from their positions are calculated. The difference in the value of a contract for a long or short position is defined as the difference between the transaction price and the current quotation multiplied by the contract unit.
That is, the change in the value of the contract is equal to the change in the price multiplied by the value of the contract. If it has a positive value, then it gives a profit to the side owning a long position in the contract, and if negative, then the profit to the party owning the short position in the contract. If the agent has several contracts, the total result is determined by multiplying by the number of contracts.

Let's describe in detail the decision-making process by the agent $i$ at the time $t = t_k$, $k = \overline{1, f-1}$:
We use the following notation:
$x_j^t$ – this is the futures price of the j-th contract at the time $t = t_k$,

$k = \overline{1, f-1}$;

$K_i^t$ – free $i$ - th agent capital by the time $t$;

$y_j^t$ – the cash price of the commodity traded in the j-th contract at the time $t = t_k$, $k = \overline{1, f-1}$;

$q_j$ – the quantity of the goods established in the j-th contract;

$m_j$ – The margin established for the j-th contract;

$p_j$ – commission fee, established for the j-th contract;

$s_r$ – the number of futures contracts concluded at the r-th step,

$s_r \leq s$, $r = \overline{1, T-1}$;

$W_i^t$ – income of the i-th agent at time $t = t_k$, $k = \overline{1, f-1}$.

**1.** At time $t = t_1$ the agent $i$ owns the starting capital $K_i^1$. He decides on the conclusion of futures contracts by number $s_1$, while making costs in the form of commission $p_j$ and margin $m_j$, $j = \overline{1, s_1}$. It is assumed that the restriction is fulfilled: $0 < \sum_{j=1}^{s_1}(m_j + p_j) \leq K_j^1$.

By the end of the first period, his income is

$$W_i^1 = K_j^1 - \sum_{j=1}^{s_1}(m_j + p_j).$$

**2.** At time $t = t_2$ the amount of free capital changes and becomes equal to $K_i^2 = W_i^1$. Futures that have been concluded earlier, already expire and bring in revenue $\sum_{j=1}^{s_1}(m_j + \lambda_j(x_j^2 - x_j^1)q_j)$, after which the agent decides to enter new futures contracts in the amount $s_2$, while making the costs of their conclusion $\sum_{j=1}^{s_2}(m_j + p_j)$. By the end of the second period, his income is

$$W_i^2 = K_i^2 + \sum_{j=1}^{s_1}(m_j + \lambda_j(x_j^2 - x_j^1)q_j) - \sum_{j=1}^{s_2}(m_j + p_j),$$

where $q_j$ - is the value of the contract, $\lambda_j = \{1, -1\}$ – characterizes the purchase or sale.

In this case, the following restriction is assumed: $W_i^2 \geq 0$.

**p.** At time $t = t_p$ the amount of free capital will change and will be equal to $K_i^p = W_i^{p-1}$. Contracts that were concluded at time p –1, will expire and yield revenue $\sum_{j=1}^{s_{p-1}}(m_j + \lambda_j(x_j^p - x_j^{p-1})q_j)$, then the agent $i$ decides to enter new futures contracts with a number $s_p$, while generating costs in the form of commissions and margins. His income by the end of period p will be equal to

$$W_i^p = K_i^p + \sum_{j=1}^{s_{p-1}}(m_j + \lambda_j(x_j^p - x_j^{p-1})q_j) - \sum_{j=1}^{s_p}(m_j + p_j).$$

In this case, the following restriction is assumed: $W_i^p \geq 0$.
And so on.

**T**. At the last moment of time $t$ = T the amount of free capital will change and become equal to $K_i^T = W_i^{T-1}$. Contracts that were concluded at the time $t$ = T-1, expire and bring in revenue

T. = T, =. = T-1

$$\sum_{j=1}^{s_{T-1}}(m_j + \lambda_j(x_j^T - x_j^{T-1})q_j).$$

At the time $t$ = T new contracts are not concluded, so the agent's income by the end of the last period will amount to

$$W_i^T = K_i^T + \sum_{j=1}^{s_{T-1}}(m_j + \lambda_j(x_j^T - x_j^{T-1})q_j).$$

It is this magnitude that should maximize each agent.

Note: At the beginning of each period, each agent owns a certain amount of free capital. Under free capital we will assume the cash available at the moment at the agent's disposal and the real goods that the agent owns for this period, expressed in money terms, in accordance with the prices of the cash market at a given time.

## 3 Description of the problem in terms of dynamic programming. Deterministic case.

As a method of optimization, we will use the method of dynamic programming.

Let's consider the case when there is no uncertainty, that is, when all the data is determined exactly and the transition of the system from one state to another is carried out with probability equal to one.

Agents make decisions at discrete instants of time $t = t_k$, k $= \overline{1, f-1}$. Therefore, we can break this process into $f-1$ a stage.
Let us take as the initial state of the system the free capital of the i-th agent at the time: $t = 0$: $K_i^0$, $i = \overline{1, n}$. The free capital of agents at the end of each period $K_i^1, K_i^2, ..., K_i^T, W_i^T$, $i = \overline{1,n}$, will be treated as system states at discrete moments of time $t = t_k$, $k = \overline{1, f-1}$.

The efficiency of the process will be characterized by income

$$W_i^p = K_i^p + \sum_{j=1}^{s_{p-1}}(m_j + \lambda_j(x_j^p - x_j^{p-1})q_j) - \sum_{j=1}^{s_p}(m_j + p_j),$$

received by each agent $i$, $i = \overline{1,n}$ for p first periods of time. Efficiency is expressed by a function

$$S_i^t(U^{t-1}) = W_i^t(U^{t-1}) - K_i^0, \; t = \overline{2,T},$$

We will maximize this function.

We introduce the function $\Re(t_k, S_i^{t_{k-1}}(U^{t_{r-2}}))$, denoting the income received by the agent for k first steps under the optimal policy. The functions $\Re(t_k, S_i^{t_{k-1}}(U^{t_{r-2}}))$, $k = \overline{1, f-1}$ satisfy the functional equations of the form

$$\Re(t_k, S_i^{t_{k-1}}(U^{t_{r-2}})) = \max_{u^{tk} \in U^{tk}} \{\Re(t_{k-1}, S_i^{t_{k-2}}(U^{t_{k-3}})) + \Delta S_i^{t_k}\},$$

Where
$\Delta S_i^{t_k} = S_i^{t_k} - S_i^{t_{k-1}}$ —

an increase in the income of the agent over a period of time $[t_k, t_{k-1}]$, and the set $U^{t_k}$ – is the set of admissible controls at this step, which is determined by the following restrictions $S_i^t(U^{t-1}) \geq 0$.

The equations given above are the basic functional equations of dynamic programming.

Let at the initial instant of time the total income of the agent be

$W_i^0 = S_i^0$,

consequently, the functions $\Re(t_k, S_i^{t_{k-1}}(U^{t_{r-2}}))$, for $k = \overline{1, f-1}$ assume the form

$$\Re(t_k, S_i^{t_{k-1}}(U^{t_{r-2}})) = \max_{u^{tk} \in U^{tk}} W_i^{t_k}.$$

As a result of applying the dynamic programming method, knowing the values of the initial capital, we obtain a sequence of functions: $\{\Re(t_k, S_i^{t_{k-1}}(U^{t_{r-2}}))\}$ – the function of the maximum income, $i = \overline{1,n}$, $k = \overline{1, f-1}$, and the corresponding optimal controls

$u_i^t = (u_{1i}^t, u_{2i}^t, ..., u_{s_t i}^t)$, $i = \overline{1,n}$, $t \in [0,T]$.

# 4 Stochastic case. Description of the game

A multi-step, non-cooperative game of n persons is being constructed

$\Gamma = <I = \{1,\ldots,n\}, \{U_i\}_{i=1}^n, \{W_i\}_{i=1}^n>$,

where $I = \{1,\ldots,n\}$ - is the set of agents,
$U_i$ – a set of strategies of the agent $i$,

$W_i : U = \prod_i U_i \to R_1$ - is the $i$ agent's payoff function, $i = \overline{1,n}$.

To solve the problem, we will use the principle of a compromise solution.

<u>Definition:</u> Let be $X$ — the set of profiles in a game $\Gamma$,

$W_i : X = U = \prod_i U_i \to R_1$, $i = \overline{1,n}$ — the payoff function of agents,

$$M_i = \max_{x \in X} W_i(x).$$

Then the compromise set is defined as follows:

$$C_H = \{\overline{x} \in X : \max_i (M_i - W_i(\overline{x})) \leq \max_i (M_i - W_i(x)), \forall x \in X\}.$$

The algorithm for finding a compromise set:
1) Calculate the winning functions for all players from each game profile.
2) Find the maximum value of the win of each player for all game profiles:

$$M_i = \max_{x \in X} W_i(x), i = \overline{1,n}.$$

3) For each profile $x \in X$ calculate the deviation of the payoff function of the i-th agent $W_i(x)$, from the maximum $M_i$, $i = \overline{1,n}$.

4) For each profile x from X, we find the maximum deviation of the difference

$$M_i - W_i(x), i = \overline{1,n},$$

that is, we compute

$$\max_i (M_i - W_i(x)).$$

5) On the set of profiles X we find a point $\overline{x}$, that delivers a minimum to the expression $\max_i (M_i - W_i(x))$, that is, we find the profile $\overline{x}$:

$$\min_{x \in X} \max_i (M_i - W_i(x)) = M_i - W_i(\overline{x}).$$

The profile in which the minimum is reached, and will be a compromise point for all players.

Since all compromise profiles are equivalent in the sense of this principle of optimality, that is, each of them gives the guaranteed win of the least satisfied player, then we will choose the strategy that corresponds to the largest of the winning functions.

Thus, provided that all agents have chosen specific controls $U^t$, the guaranteed income of the agent $i$, at the last step is equal to the value

$$K_i^T + \min_{x \in X} \max_i (\sum_{j=1}^{s_{T-1}} (m_j + \lambda_j (x_j^T(U^{T-1}) - x_j^{T-1}(U^{T-2}))q_j)).$$

Using the method of dynamic programming, the guaranteed income at the final point in time with fixed member $U^t$ control is calculated. Agent $U^t$ control at each step is selected in accordance with the principle of a compromise solution, then at each step a single control is selected by the method of backward procedures, resulting in a single process trajectory.

## 5 Acknowledgements

The work is partly supported by work RFBR No. 18-01-00796.